\documentclass[12pt,preprint]{aastex}

\lefthead{Bekki}
\righthead{Two-faced dwarf irregulars}

\begin{document}

\title{Origin of dwarf irregular galaxies with outer early-type structures}

\author{Kenji Bekki} 
\affil{
School of Physics, University of New South Wales, Sydney 2052, Australia}

\begin{abstract}

Recent observations have reported
that some  gas-rich dwarf irregular (dIrr) galaxies appear to have
spherical distributions in the outer  underlying  old and
intermediate-age
stellar populations (e.g., NGC 6822).
These observations imply that some dIrr's have 
two distinct (or ``two-component'')
structures, i.e., inner disky and outer
spherical ones,
though the number fraction of dIrr's with such structures
remains observationally unclear.
We discuss how such two distinct   structures are formed
during dIrr formation based on observations and simulations.
Our numerical simulations  show that 
the remnants of  mergers between two gas-rich dIrr's
with initially extended gas disks 
can have both extended spheroids composed of  older stellar populations
and  disks composed mostly of gas and young stars.
The simulated remnants with two distinct structures
can be still identified as dIrr's owing to
the presence of star-forming regions.
The structural  properties of outer spherical structures in
dIrr's formed from dIrr-dIrr merging  depend on initial conditions
of merging,
which suggests that outer structures in dIrr's can be diverse.
We also discuss other possible physical mechanisms for the formation
of outer spherical structures composed of older stars in dIrr's. 

\end{abstract}


\section{Introduction}

Luminous galaxies like the Milky Way 
can be formed from merging and accretion of
numerous low-mass galactic building blocks
like dwarf galaxies in hierarchical  galaxy formation
scenarios  based on the cold dark matter cosmology 
(e.g., White \& Rees 1978).
Physical properties of dwarf galaxies
thus have been discussed extensively in many different contexts
of galaxy formation and evolution,
such as 
physical mechanism for
galaxy-scale star formation process in dIrr's (e.g., Hunter 1997)
and evolutionary links between different types of dwarfs (e.g., Grebel et al. 2003).
Stellar populations, structures and kinematics of HI gas,
and spatial distributions of young stars
in gas-rich dIrr's  particularly have attracted 
much attention from theoretical works on formation of stars and
star clusters (e.g., Gallagher \& Hunter 1984;  Hunter 1997)
and inner structures of their dark matter halos (e.g., Burkert 1995).

Although previous observations revealed 
fundamental physical properties of star-forming regions, young stellar
populations,  and HI gas in dIrr's (e.g., Hunter 1997),
structural and kinematical properties of {\it underlying older stellar
populations}  remain fairly unclear.
Observational studies of dIrr's
based on broadband imaging ($UBVJHK$)  
revealed that the structures of main disk components in dIrr's 
can be described as exponential profiles (Hunter \& Elmegreen  2006).
Furthermore observational studies on   mid-infrared properties of dIrr's
suggested that the  overall structure of a dIrr is the same 
in different passbands for most dIrr's 
(Hunter et al. 2006).
Since these observations are  based on integrated stellar lights,
they did not allow the authors to
investigate {\it separately}  structures of stellar populations with different
ages in detail.

Recent observational studies on spatial distributions
of individual older (e.g. RGB stars) and intermediae-age (e.g., AGB ones)
stars in nearby dIrr's  have shown
that some dIrr's 
(e.g., NGC 6822 and the Small Magellanic Cloud; SMC)
have  apparently  spherical distributions
in the outer underlying older stellar populations
(e.g., Battinelli et al.  2006;  Cioni et al. 2006; Demers et al. 2006).
These observations combined with the above-mentioned
imaging ones (e.g., Hunter et al. 2006) imply that some dIrr's possibly 
have inner main disky structures
with younger stellar populations
surrounded by extended spherical structures composed of older stellar 
populations: some dIrr's have two distinct (or ``two-component'') structures.
It is, however, unclear how such two-component structures were formed
in some dIrr's (e.g., NGC 6822).

The purpose of this {\it Letter} is thus the following two.
The first purpose is to introduce a
classification scheme that divides
 structures of outer underlying older stellar populations
in dIrr's into the following two:
``early-type'' (spherical) and ``late-type'' (disky).
The possible candidates of dIrr's with early- and late-types outer structures
are summarized in Table 1
and the schematic diagram for
the proposed structure classification
scheme is shown in Fig. 1.
Outer early-type structures in NGC 6822, IC 10, and IC 1613,
are confirmed by  Battinelli et al. (2006), Demers et al. (2004),
and  Battinelli et al. (2007),  respectively.

Tikhonov (2006) found that young stars in a dIrr form a thin disk with the size
corresponding to its visual size whereas older ones form a extended thick disk
for a number of dIrr's 
(e.g., DDO 216, ESO 381-018, IC 1574, and IC 3104):
these  dIrr's with outer late-type  structures  
may well be regarded as normal dIrr's. 
Furthermore Demers et al. (2003) found that IC 1613 does not have an outer
spherical halo composed of old stars (e.g., red giants)
but shows a disky structure  composed of carbon stars: IC 1613 is an example
of dIrr's with late-type structures.

The second purpose of this paper is to
demonstrate that
gas-rich dwarf-dwarf merging 
can transform dIrr's with outer late-type  structure
into those with early-type ones.
We mainly describe physical properties of
the remnants of dwarf-dwarf mergers  with different model
parameters in order to discuss the origin of dIrr's with outer early-type structures.
It should be stressed here that dIrr's with early-type structures discussed
in the present paper are different from 
``transition type dwarfs'' (e.g., Phoenix and DDO 210)
which are intermediate-class of objects 
between  dIrr's and dEs/dSphs and have HI gas mass of at most a few 
$10^6 {\rm M}_{\odot}$ (e.g., Grebel et al. 2003): formation and evolution
of the transition-type dwarfs 
are discussed in detail by Skillman et al. (2003). 
We do not intend to discuss dIrr's in the Local Group,
though observations of them provided some clues to formation and evolution of dIrr's
in general (e.g., Mateo 1998).

Recent observational studies on stellar kinematics in dIrr's have shown that there can
be kinematical differences between stars and HI gas: Hunter et al. (2002) 
found that stellar components in NGC 4449 have a much smaller amount of rotation
in comparison with gaseous ones. Since we consider that dIrr's with
these gas-star kinematical differences are possible candidates of
those with outer early-type structures,
we list the candidates in Table 1: such star-gas  kinematical differences in NGC 6822
and the SMC are observationally confirmed by  Demers et al. (2006)
and Harris \& Zaristky (2006), respectively.

\section{The model}

We investigate chemodynamical evolution of gas-rich mergers
between dIrr's  with stellar and gaseous disks embedded in massive
dark matter halos by using our original
Nbody/SPH (TREESPH) codes
with models of chemical evolution
and globular  cluster formation  (Bekki et al. 2002).
Since the details of numerical techniques and methods (e.g., ways to model
chemical enrichment processes) are already given in above 
paper and our forthcoming papers (Bekki \& Chiba 2008),
we briefly describe them in the present paper.
We adopt the Burkert profile (Burkert 1995) for the radial
density profile of the dark matter halo of a dIrr.
The total masses of the dark matter,
the stellar  component,
and the gaseous one
for the dIrr
are set to be $8.0 \times 10^9 {\rm M}_{\odot}$,
$4.0 \times 10^8 {\rm M}_{\odot}$,
and $8.0 \times 10^8 {\rm M}_{\odot}$, respectively.

The dIrr is described as a  disk 
and the radial ($R$) and vertical ($Z$) density profile
of the initially thin  
are  assumed to be
proportional to $\exp (-R/R_{0}) $ with scale length $R_{0}$ = 1 kpc 
and to  ${\rm sech}^2 (Z/Z_{0})$ with scale length $Z_{0}$ = $0.2R_{0}$,
respectively.
The HI diameters of gas-rich galaxies are generally observed to be
significantly larger
than their optical disks (Broeils \& van Woerden 1994). 
Therefore the ratio ($s_{\rm g}$) of the stellar disk size  ($r_{\rm s}$) 
to the   gaseous one  ($r_{\rm g}$)
is assumed to be a free parameter.
We however mainly show the results of the model with $s_{\rm g}$=4.0.
Star formation in gas
is modeled by converting  the collisional
gas particles
into  collisionless new stellar particles according to 
the Schmidt law (Schmidt 1959)
with exponent $\gamma$ equal to  1.5
(e.g., Kennicutt 1998).
The stars formed from gas are called ``new stars'' 
whereas stars initially within a disk  are called ``old stars''
throughout this paper.
The simulations have mass and size resolutions of $10^3 {\rm M}_{\odot}$
and 50 pc, respectively, for stars in all models.

The mass ratio of the two merging dIrr's ($m_2$), 
the pericenter distance ($r_{\rm p}$),
and the eccentricity ($e_{\rm p}$) are 
assumed to be free parameters.
The orbit of the two dIrr's is
set to be the same as
the $xy$ plane and the distance between
the center of mass of the two dIrr's
is  20 kpc.
The spin of each galaxy in a merger
is specified by two angles $\theta_{i}$ and
$\phi_{i}$, where suffix  $i$ is used to identify each galaxy.
$\theta_{i}$ is the angle between the $z$ axis and the vector of
the angular momentum of a disk.
$\phi_{i}$ is the azimuthal angle measured from the $x$ axis to
the projection of the angular momentum vector of a disk onto the $xy$
plane.
Although we run many models with different orbits,
we mainly show models with 
$r_{\rm p}=1$ kpc, $e_{\rm p}=1.0$, 
$\theta_{1}=30^{\circ}$,
$\theta_{2}=120^{\circ}$,  $\phi_{1}=90^{\circ}$,
$\phi_{2}=30^{\circ}$.

We mainly describe  the results of the four representative models
with $m_{2}$= 0.1, 0.3, 0.5, and 1.0, 
because the merger remnants of these  models show
(i) early-type morphologies 
in  their underlying older stellar populations
and (ii) diversity in the morphologies.
We  do not intend to discuss 
star formation histories,
chemical evolution, and formation of globular clusters
in dIrr's with outer early-type structures  and their dependences on model 
parameters: these will be discussed in our forthcoming papers. 
The stellar-to-mass-to-light-ratio in the $B$-band 
($M_{\rm s}/L_{\rm B}$) is assumed to be 1.78 for old stars
and 0.25 for new ones so that we can convert
the simulated  stellar mass densities 
into the $B-$band surface brightness (${\mu}_{\rm B}$).
The adopted $M_{\rm s}/L_{\rm B}$ for old (new) stars are 
from the tables of stellar population synthesis models
with $ {\rm [Fe/H]}=-1.28 $ and ages of 5 (0.5) Gyr by 
Vazdekis et al. (1996).

\section{Results}

Fig. 2 shows that spatial distributions of old and new star are
significantly  different in the sense that spatial distributions
of new stars are much more compact and flattened than those
of old ones in the four representative models with different $m_{2}$.
Spatial distributions of underlying old stars depend on $m_2$
such that they are more spherical in the models with larger $m_2$.
Extended gas disks surrounding new stars can be clearly seen
in all of the four models
and very low-level star formation 
with the rate of $0.03-0.1 {\rm M}_{\odot}$ yr$^{-1}$  
is still ongoing within the gas disks in a sporadic manner for the
models.
The formation of extended gas disks in merger remnants
is due essentially to
the presence of extended gas disks in 
merger precursor  dIrr's.
Thus these remnants with  outer (flattened) spheroids composed
of old stars,  disky structures of young stars,
and extended gas disks
can be morphologically  classified  as dIrr's with outer early-type structures.

About 30\% of initial gas masses
corresponding to $\sim (2-5) \times 10^8 {\rm M}_{\odot}$
can be still within the remnants, which means that the remnants
can be regarded as gas-rich dwarfs.
Fig. 3 shows that the gas mass fractions ($f_{\rm g}$) are significantly  higher
in the outer parts of the merger remnant for the model with $m_{2}=0.3$
($f_{\rm g} \sim 0.6$ for $R \sim 5$ kpc).
This radial dependence of gas mass fraction can be seen in models
with different $m_{2}$, which implies that    dIrr's with early-type structures
formed from
merging of  dIrr's have most of their HI gas in their outer parts. 
Fig. 3 also shows that although there is a strong
concentration of new stars in the  central region ($R\sim 1$) kpc of the merger
remnant ($f_{\rm ns} >0.5$), the dark matter halo already
dominates in mass there 
(i.e., $f_{\rm b} <0.3$).

Fig. 4 shows that $B-$band surface brightness distributions 
(${\mu}_{\rm B}$) estimated separately for old and new stars
in the merger remnant with $m_{\rm 2}=0.3$.
New stars have significantly  higher ${\mu}_{\rm B}$
than old ones within the central 5 kpc of the remnant  so that
the optical morphology of the remnant can be  determined largely by
the distribution of new stars:
in spite of its regular 
(flattened) spherical  morphology  
in the underlying old stars,
this remnant can be still
classified as a dIrr  rather than a dSph (or dE)
in the canonical morphological classification scheme,
because  the outer spherical distribution  of old stars ($R>1$ kpc) with
very low ${\mu}_{\rm B}$ ($>27$ mag arcsec$^{-2}$)
is hard to be detected.
It is confirmed that
low surface brightness outer envelopes 
composed mostly of old stars are common in
the remnants of dwarf-dwarf mergers with different $m_2$
in the present study.
Thus the present study demonstrates that two dIrr's 
with late-type structures can be
transformed into one dIrr with an outer early-type structure through gas-rich merging.

\section{Discussion and conclusions}

The present study has shown that if gas-rich dIrr's have extended 
gas disks, 
 merging between the two can hardly transform them into one gas-free
dSph's owing to the presence of the gas disk and star-forming regions in the remnant.
The Faint Irregular Galaxies GMRT Survey ({\it FIGGS}) has recently revealed
that the medial ratio of $r_{\rm g}/r_{\rm s}$ is about 2.4 for dIrr's with
a median $M_{\rm B} \sim -13$ mag (Begum et al. 2008).
This observation combined with the present results
implies  that morphological transformation from two gas-rich dIrr's into
gas-free dSph's via merging is highly unlikely, at least, in the present universe.
We thus suggest that the remnants of merging between dIrr's 
can be identified as dIrr's with outer early-type structures and gas  such as NGC 6822
(Demers et al. 2006).
Possibly, stripping of HI gas due to tidal interaction with
luminous galaxies and ram pressure of hot gas in group and cluster
environments could drive further evolution from
dIrr's with early-type structures into gas-free dSph's.

Given that merging of low-mass galactic building blocks can be  a fundamental
mode of galaxy formation in the  hierarchical clustering scenarios 
(e.g., White \& Rees 1978),
the present dIrr's with outer early-type structures
could be the relic of such hierarchical galaxy formation. 
However, it is observationally unclear what fraction of dIrr's
are actually those with early-type structures
owing to the lack of extensive statistical studies
on spatial distributions of older stellar populations in the outer 
parts of dIrr's.
van den Bergh (1988) suggested that typical dIrr's are triaxial systems
with axis ratio of 1.0:0.9:0.4 based on the observed distributions of inclination
angles of nearby dIrr's. 
Observational studies on the distribution of ellipticities in 
light distributions of 30 low-surface
brightness dIrr's suggested that the intrinsic shape of the dIrr's is triaxial
and slightly less spherical than dEs (Sung et al. 1998).
Although these observations imply that inner main components  of dIrr's  have 
more flattened distributions,
it remains unclear whether outer older ones also have such distributions. 

As shown in Fig. 4,
the outer underlying structure composed of older stars in a dIrr
formed by merging
is hard to be detected owing to its very low surface brightness
(e.g., ${\mu}_{\rm B} > 30$ mag arcsec$^{-2}$ for $R>3$ kpc),
even if the mass fraction of the older stars is significant.
This suggests that observations based on spatial  distributions
of individual bright stars  (e.g., RGB/AGB/carbon stars)
rather than on broadband imaging would be better to reveal
the presence of outer extended structures composed mostly of
older stars in dIrr's. 
Although the presence of intermediate-age stars within the outer halos
for a number
of dIrr's has been reported (e.g., Albert et al. 2000),
the details of their spatial distributions are not so clear for
some of the dIrr's.
As suggested in the present simulations,
the possible presence of two distinct structures (i.e., inner disks and outer
spheroids) would have valuable information on the  formation of dIrr's.
Thus it is doubtlessly worth  while for future observations
to detect outer spherical or disky structures composed of
older stars in dIrr's.

Although nearby dIrr's are observed to show signs of rotational kinematics
in their {\it gaseous components} (e.g., Mateo 1998 for a review),
it is observationally  unclear whether their stellar components 
have also rotational kinematics (e.g., Hunter et al. 2002; 2005).
The present study predicts  that $V/\sigma$
can be significantly  different between gas and stars in the merger remnants
in the sense that
their  stellar $V/\sigma$ are smaller than gaseous ones.
Therefore,  gas-rich dIrr's with kinematically hot
stellar components are promising candidates
of the remnants of mergers between dIrr's with
extended gas disks.
Thus future  systematical studies of stellar kinematics
of dIrr's will  help us to understand what fraction of dIrr's
are those with early-type structures  and thus possibly formed from dwarf-dwarf merging.

Both the presence  of intermediate-age stars in very outer regions of dIrr's 
(e.g., Albert et al. 2000; Letarte et al. 2002)
and apparently spherical distributions of these stars (e.g., Battinelli et al.  2006)
can be explained by the present merger scenario of dIrr formation, if merging
can happen  $2-10$  Gyr ago.
However, dIrr-dIrr merging can be only one of several  possible physical mechanisms
for the formation of outer structures composed of intermediate-age stars.
For example,
minor merging of very tiny dwarfs with intermediate-age stars onto a dIrr
can be also responsible for the presence of AGB and carbon stars in the outer region
of the dIrr, though such minor merging can hardly produce an outer  spherical
structure. 
If supersonic  gaseous outflow  driven  by  central star-formation activity
in a dIrr can interact with the outer halo gas and 
consequently trigger star formation  in the high-density
shocked gaseous regions,  such new stars might well be identified as intermediate-age
halo stars
several Gyrs later: dynamical relaxation after the formation of halo stars
can be responsible for the formation of a spherical structure
around the dIrr  in this scenario.
We thus suggest that physical properties of older and intermediate-age stars
in the very outer regions of dIrr's can provide valuable
information on formation and evolution of dIrr's.

\acknowledgments
I (K.B.) am   grateful to the anonymous referee for valuable comments,
which contribute to improve the present paper.
K.B. acknowledges the financial support of the Australian Research
Council throughout the course of this work.
The numerical simulations reported here were carried out on
SGI supercomputers
at Australian Partnership for Advanced Computing (APAC) in Australia
for our  research project.

\begin{deluxetable}{cc}
\footnotesize
\tablecaption{Possible candidates of 
dIrr's with outer early-type (i.e., spherical)  structures,
outer late-type (i.e., disky) ones, and 
kinematical differences between stars and HI gas.  \label{tbl-1}}
\tablewidth{0pt}
\tablehead{
\colhead{dIrr properties} & \colhead{Examples}  }
\startdata
Outer spherical structures &   NGC 6822, IC 10, IC 1613    \\
Outer disky structures  &   DDO 216, ESO 381-018, IC 1574   \\
Star-gas kinematic differences  &  NGC 6822, SMC, NGC 4449  \\
\enddata
\end{deluxetable}

\clearpage

\begin{figure}
\plotone{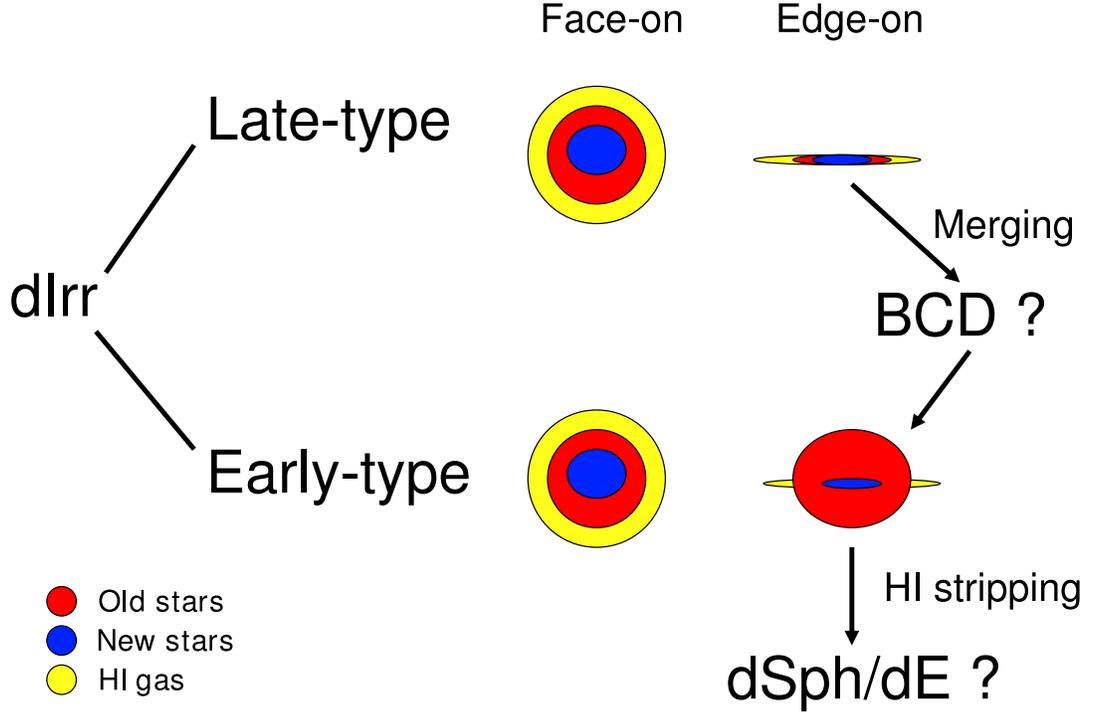}
\figcaption{
A schematic diagram for a possible diversity in
outer underlying structures of older stellar populations (e.g., RGB/AGB stars)
in dIrr's. 
Here  dIrr's are divided  into  (i) those with outer ``early-type'' 
(i.e., spherical) structures and (ii) those with ``late-type'"
(i.e., disky) ones according to the outer underlying structures of older 
stellar populations.
Old stars, new (or young) ones, and HI gas are shown in red, blue,
and yellow, respectively,
both for the face-on and the edge-on views.
dIrr's with late-type structures (e.g., DDO 216) are basically dwarfs with rotating
stellar and gaseous disks whereas those with outer early-type
structures  (e.g., NGC 6822) have both
spherical structures  of older stars  and
disky ones  of young and (old) stars and HI gas.
The degree of flattening can be different for outer spheroids
and the spheroids can be supported  dynamically by
velocity dispersion and  by rotation.
Possible stellar halos
composed of very old stars
(e.g. Minniti \& Zijlstra 1996; Aparicio \&
Tikhonov 2000) are not used for this morphological classification.
It should be here stressed that owing to the presence 
of intermediate-age stars,  the outer spherical structures
are different from stellar halos dominated by very old stars with ages
larger than 10 Gyrs in luminous galaxies like the Galaxy.
dIrr's with outer early-type structures are different from ``transition type''
(dSph/dIrr) dwarfs owing to the presence of star-forming 
regions and abundant  HI gas.
One of possible evolutionary links between 
dIrr's with late-type and early-type structures
is shown: dIrr's with late-type structures can evolve into 
blue compact dwarfs (BCDs) and further into dIrr's with early-type ones
by dwarf-dwarf merging (Bekki  2008).
Complete or partial stripping of HI gas via some physical processes
(e.g.,  tidal interaction with other galaxies
or ram pressure) would transform dIrr's with early-type structures into 
dIrr/dSph transition dwarfs and gas-free
dSph's (or dE's).
\label{fig-1}}
\end{figure}

\begin{figure}
\plotone{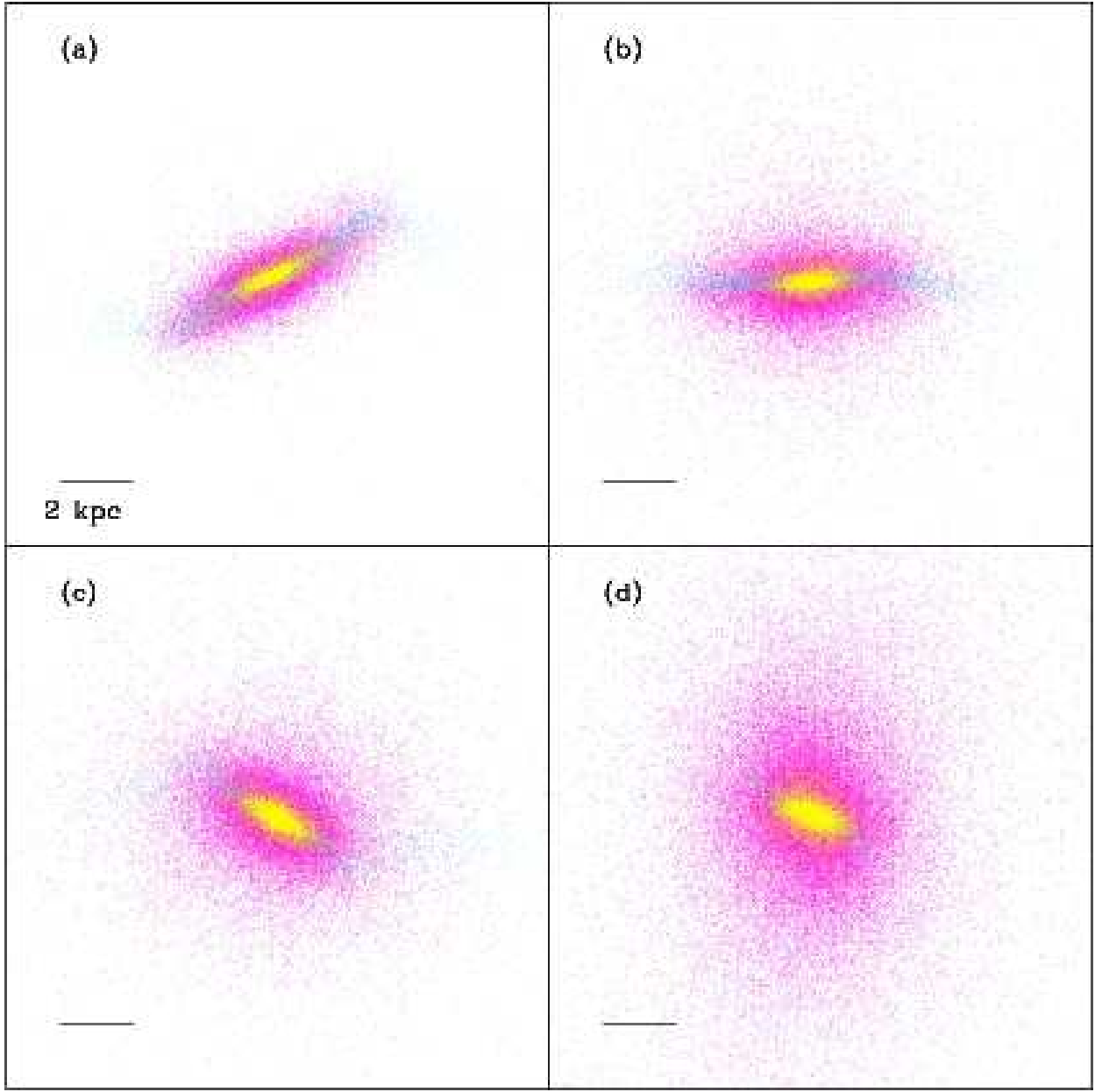}
\figcaption{
Final mass distributions of old stars (magenta), gas (cyan),
and new stars (yellow) in
the remnants of dwarf-dwarf mergers for
the four representative models
with $m_{\rm 2}$=0.1 (a), 0.3 (b), 0.5 (c), and 1.0 (d).
A bar in the lower left corner in each panel measures 2 kpc.
It is clear that the remnants have inner  disky 
(or very flattened spherical) structures 
of new stars,  outer (flattened) spherical ones of old stars,
and extended HI gas disks surrounding new stars.
These remnants can be classified dIrr's with outer early-type structures.
\label{fig-2}}
\end{figure}
 
\begin{figure}
\epsscale{0.7}
\plotone{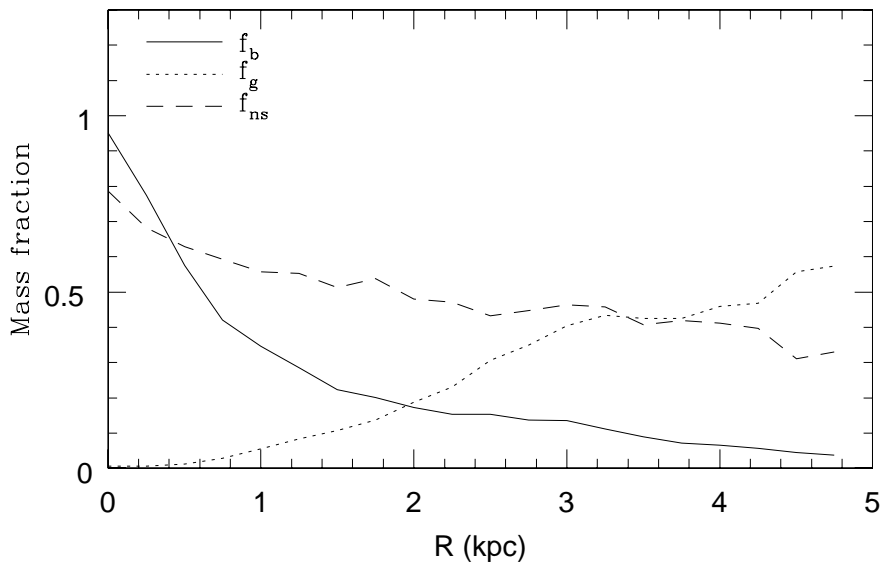}
\figcaption{
The radial dependences of mass fractions in the merger remnant
for the model with $m_{\rm 2}=0.3$:
$f_{\rm b}$ is the mass fraction of baryon (gas and stars)
to all components (solid), $f_{\rm g}$ is that  of gas
to baryonic components (dotted), and $f_{\rm ns}$ is that of 
new stars to all stellar components  (dashed).
\label{fig-3}}
\end{figure}

\begin{figure}
\epsscale{0.7}
\plotone{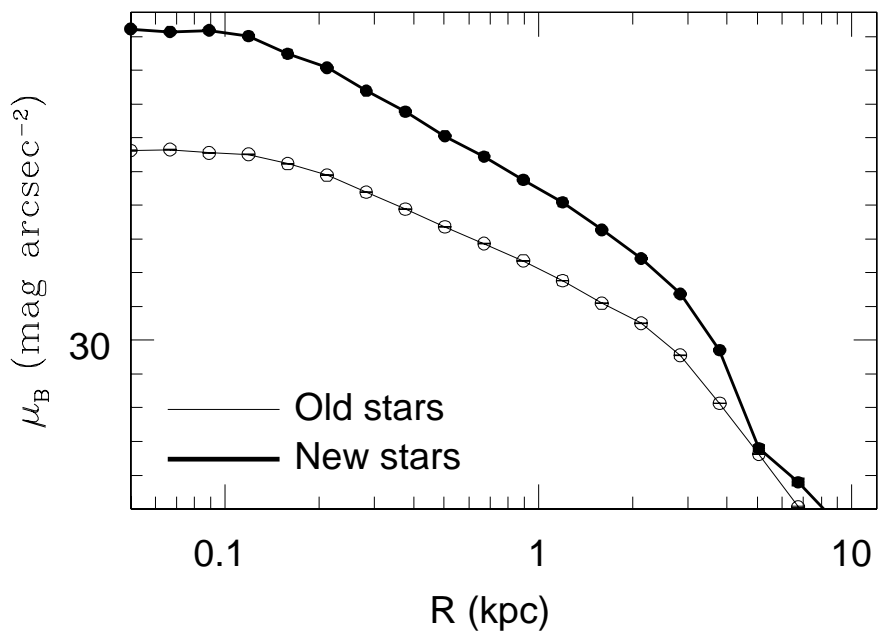}
\figcaption{
The $B$-band surface brightness profiles (${\mu}_{\rm B}$)
for old stars (thin)
and new ones (thick) for the merger remnant in the model with $m_2=0.3$.
\label{fig-4}}
\end{figure}

\end{document}